\documentclass[%
reprint,
amsmath,amssymb,
aps,
]{revtex4-2}

\usepackage{graphicx}
\usepackage{dcolumn}
\usepackage{bm}
\usepackage{xcolor} 

\begin{document}

\title{LIDAR with Optically Amplified Thermal Light}

\author{Darren Ming Zhi Koh$^{1}$}
\author{Peng Kian Tan$^{1}$}
\email{pengkian@gmail.com}
\author{Christian Kurtsiefer$^{1,2}$}
\email{christian.kurtsiefer@gmail.com}
\affiliation{$^{1}$Centre for Quantum Technologies, 3 Science Drive 2, Singapore 117543}
\affiliation{$^{2}$Department of Physics, National University of Singapore, 2 Science Drive 3, Singapore, 117551}

\date{\today}

\begin{abstract}
Thermal light exhibits intensity fluctuations or photon bunching that has been used as a source of entropy or temporally correlated light. However, its low spectral density limits its use in practical applications like LIDAR. One method of remedying this is optical amplification to boost its brightness. In this paper, we show that the normalized second order coherence function $g^{(2)}(\tau)$ of a thermal light source is preserved after amplification. We then use amplified thermal light to demonstrate a LIDAR raster scan of a building at 1.3\,km with a depth resolution of 2\,cm.
\end{abstract}


\maketitle

\section{Thermal Light Sources}
Thermal light, such as blackbody radiation or spontaneous emission from an ensemble of atoms, can be described as a superposition of electric fields with no fixed phase relationship \cite{loudon}. This gives rise to intensity fluctuation on a timescale corresponding to its coherence time. These fluctuations have been used as a source of entropy for probabilistic photonic computing \cite{bruckerhoff2024probabilistic} as well as in quantum random number generation (QRNG) \cite{williams2010fast,guo202140}. The associated correlation, often referred to as photon bunching, is also desirable for LIDAR applications as it does not require additional modulation, has no range limitation from periodicity, and is resistant to turbulence \cite{wagner2021intensity,tan2016optical}. It has been shown to work at long distances with cm depth resolution, and can achieve spatial resolution 14 times better than the diffraction limit \cite{zhu2012thermal,tan2023practical,liu2025active}. Thermal light sources are also used in many other applications such as optical time domain reflectometry \cite{han2023ultra}, ghost imaging \cite{valencia2005two,zhang2005correlated}, ghost spectroscopy \cite{janassek2018ghost}, clock synchronization \cite{peh2026clock}, and as a pump for nonlinear optical processes \cite{wakui2025modelocked,valero2021high}.

To observe the intensity fluctuations, the coherence time has to be at least on the order of 10\,ps, corresponding to the response time of state-of-the-art photodetectors \cite{shi2025ultrafast,chang2021detecting}. However, thermal light sources often have a broad frequency spectrum, resulting in coherence times on the order of femtoseconds. While this is resolvable with two-photon absorption processes \cite{boitier2009measuring}, a more practical solution is to increase the coherence time through spectral filtering at the cost of brightness \cite{tan2014measuring}.

\begin{figure}
  \centering
  \includegraphics[width=1\columnwidth]{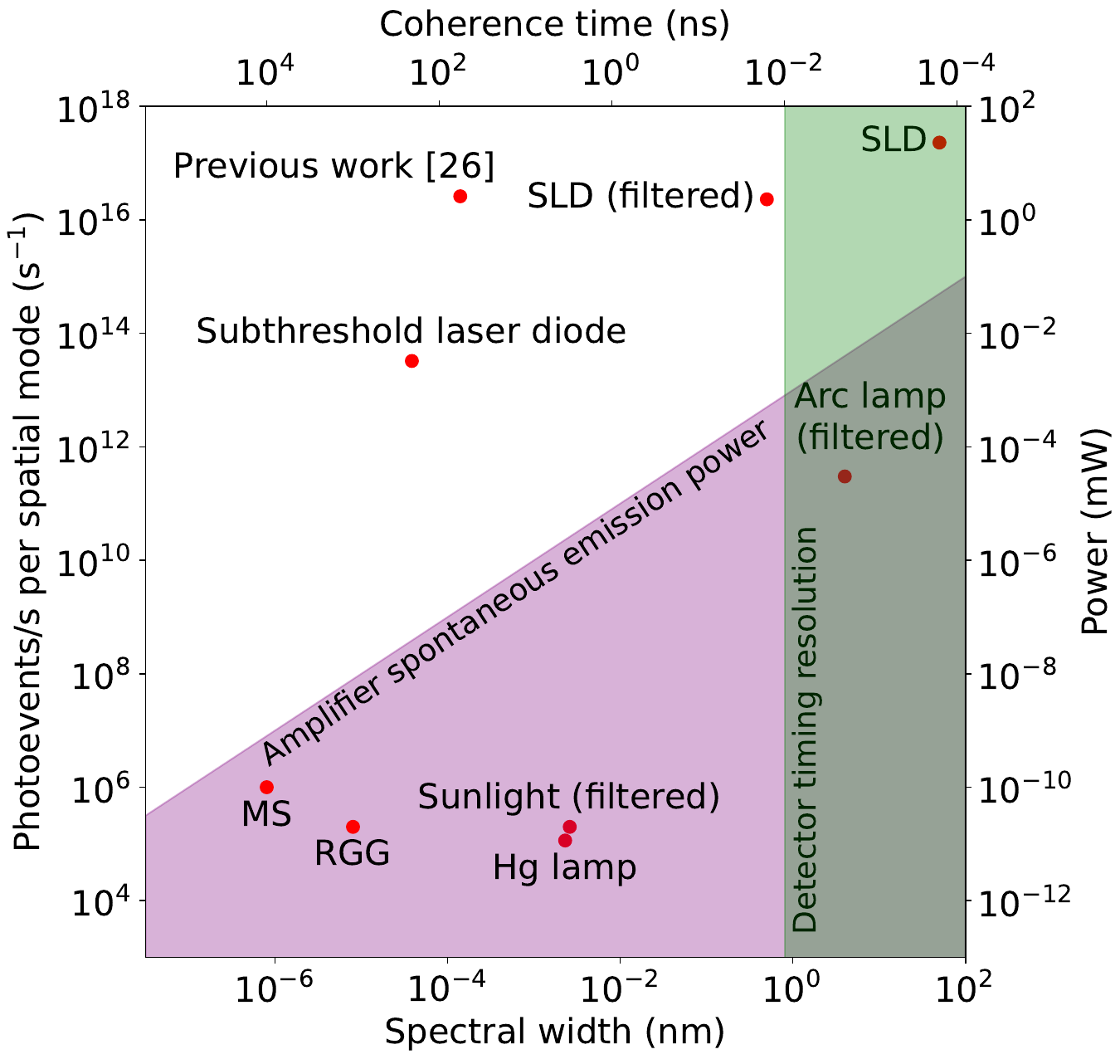}
\caption{\label{fig:source_comparison}
  Comparison of brightness and spectral widths of different thermal light sources. Light sources in the green region have unresolvable intensity fluctuations. Light sources in the purple region are dimmer than the ASE of an optical amplifier, assuming a uniform ASE spectral density of 1\,\textmu W/nm. Amplifiable light sources with resolvable coherence time are in the white region. Microsphere suspension (MS) and rotating ground glass (RGG) are pseudothermal light sources whose intensity fluctuation arise from scattering light through a time-varying dispersive medium. As such, their coherence time do not depend on their spectral width. SLD: superluminescent diode.
}
\end{figure}

Optical amplification through stimulated emission has been used to recover brightness, but this requires the optical power density to be larger than the uncorrelated amplified spontaneous emission (ASE) introduced \cite{beshr2015study,becker1999erbium}. This restricts amplification to only a few primary sources of thermal light as shown in Fig. \ref{fig:source_comparison}. Alternatively, one could also use a non-linear optical process such as optical parametric amplification for amplification, which has been demonstrated to work at the single photon level \cite{lamas2002experimental,de2008entanglement}. This could allow for dimmer primary thermal light sources.

The use of amplified thermal light has been explored in some of the applications above \cite{williams2010fast,guo202140,han2023ultra,wakui2025modelocked}. In those works, the amplified light has been shown to also exhibit intensity fluctuations or photon bunching. In this work, we investigate the cross-correlation between the intensity before and after amplification, showing that the normalized second order coherence function $g^{(2)}(\tau)$ is largely unaffected by amplification. We then make use of amplified thermal light to perform a LIDAR raster scan of a building at a distance of 1.3\,km with 2\,cm depth resolution.

\section{Intensity correlations after amplification}
The normalized second order coherence function $g^{(2)}(\tau)$ is given by
\begin{align}
	g^{(2)}(\tau)=\frac{\langle I(t+\tau)I(t)\rangle}{\langle I(t)\rangle^2},
\end{align}
where $I(t)$ is the intensity of the light. Thermal light features a photon bunching peak at $\tau=0$ that decays on a characteristic timescale, the coherence time of the light. The exact form can be determined from its power spectral density $S(f)$ using the Wiener-Khinchin theorem and the Siegert relation \cite{loudon,siegert1943fluctuations},
\begin{align}
	g^{(1)}(\tau)&\propto \mathcal{F}^{-1}\left[S(f)\right],\label{eqn:Wiener-Khinchin theorem}\\
	g^{(2)}(\tau)&=1+\left|g^{(1)}(\tau)\right|^2,\label{eqn:Siegert relation}
\end{align}
where $g^{(1)}(\tau)$ is the normalized first order coherence function.

\begin{figure}
  \centering
  \includegraphics[width=1\columnwidth]{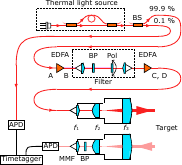}
  \caption{\label{fig:setup}
    Experimental setup comprising mostly of single mode fiber based components. Separate telescopes are used for illuminating and collecting light off the target. APD: single photon avalanche photodiode, BS: beamsplitter, EDFA: erbium-doped fiber amplifier, Pol: polarizer, BP: bandpass filter, MMF: multimode optical fiber.}
\end{figure}

The light source used in the experiment is described in \cite{yeo2026nearlosslessmethodgeneratingthermal} and was chosen for its high brightness and comparatively long coherence time. Laser light is injected into a series of $n$ cascaded unbalanced Mach-Zehnder interferometers. The path difference in each interferometer is chosen to be larger than the coherence length of the laser light such that the output field is a superposition of phase-independent fields. This results in 
\begin{align}
	g^{(2)}(\tau)&=1+\left(1-\frac{1}{2^n} \right)\left|g^{(1)}(\tau)\right|^2,
\end{align}
which for $n\rightarrow\infty$ reduces to the Siegert relation (Eqn. \ref{eqn:Siegert relation}). This is because when $n\rightarrow\infty$, the output field is exactly the model of thermal light described in the introduction.

A schematic of the experimental setup is shown in Fig. \ref{fig:setup}. The primary light source consists of a distributed feedback laser injected into only one interferometer ($n=1$). The laser has a Lorentzian spectral profile centered at wavelength 1550\,nm, exhibiting $g^{(1)}(\tau)\propto e^{-\left|\tau\right|/\tau_c}$ \cite{henry1982theory}. Thus,
\begin{align}
	g^{(2)}(\tau)&=1+\frac{1}{2}e^{-2\left|\tau\right|/\tau_c},
\end{align}
where $\tau_c$ is the coherence time of the laser light.

The interferometer is constructed from single mode fiber based components to ensure good spatial mode overlap at the beamsplitters. A spool of optical fiber (SMF-28) of length 20\,km serves as the delay arm of the interferometer. The choice of delay length is limited by the available fibers in the lab, but a shorter delay fiber can also be used depending on the laser light coherence length. A fiber paddle (not shown) in the other arm is used to match the polarization of both arms at the output beamsplitter.

The light is split with a 99.9:0.1 fiber beamsplitter, with 0.1\% of the light sent to a reference detector, and the rest to the amplifier chain.

\begin{figure}
  \centering
  \includegraphics[width=\columnwidth]{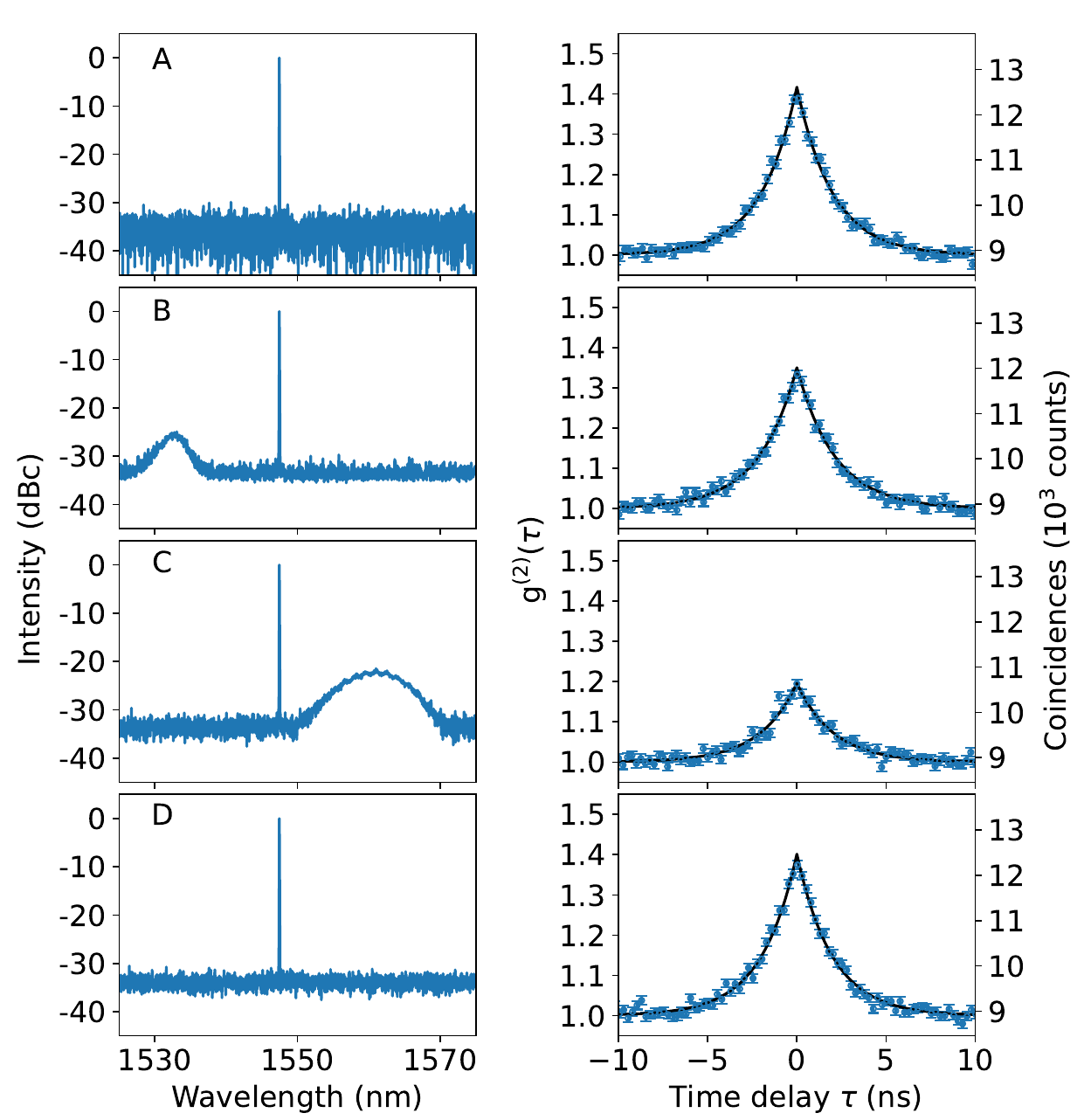}
  \caption{\label{fig:spectrum and g2}
Comparison of optical spectra and $g^{(2)}(\tau)$ at different stages of amplification. The spectral resolution is limited to 4\,GHz by the spectrometer. Broad peaks at 1530\,nm and 1560\,nm at B and C are due to the uncorrelated ASE from the EDFA. The extracted $g^{(2)}(\tau)$ has a floor at $g^{(2)}=1$ corresponding to accidental coincidences of 8900\,counts per time bin. Error bars are based on  Poissonian counting statistics of $\sqrt{N}$. \textbf{First row} No amplification, point A in the setup. \textbf{Second row} One EDFA, point B in the setup. \textbf{Third row} Both EDFAs, point C in the setup. \textbf{Fourth row} Both EDFAs with filtering of ASE, point D in the setup.}
\end{figure}

Amplification was done using a pair of erbium-doped fiber amplifiers (EDFA, Thorlabs EDFA100s, Simtrum STEDFA-C-HP-BA-40-SM-B). Amplification introduces ASE \cite{becker1999erbium}, which is uncorrelated to the input light. This degrades the correlation between the input and output light. Most of the ASE is spectrally distinct from the input light (see Fig. \ref{fig:spectrum and g2}), and was removed using a band pass filter with 4\,nm passband and a polarizer.

The light is then sent to a second detector at either points A, B, C, or D. Measurements at A, B, and C were done without the filter stage between the EDFAs.

We used a pair of actively quenched indium gallium arsenide (InGaAs) single photon avalanche photodiodes (APD, IDquantique ID Qube LN), with a quantum efficiency of 10\,\% and a dead time of 5\,\textmu s and 2\,\textmu s, respectively. These settings were chosen to minimize dark counts and afterpulsing, and to maximize the maximum count rate.

A timetagger (S15 instruments TDC2) with 3.6\,ps nominal resolution was used to record the photoevent timestamps. The combined instrument response function (IRF) of the detectors and timetagger was characterized using a spontaneous parametric downconversion photon pair source with the technique described in \cite{shen2021wide}. This resulted in an IRF timing jitter $\tau_j\approx500$\,ps FWHM.

Photoevents of the two detectors were recorded with an integration time $T$=10\,min. We then compute the intensity cross correlation by histogramming the time difference between photoevents at the two detectors with a time bin width $\delta_t$=0.25\,ns. To extract $g^{(2)}(\tau)$, we normalize the histogram with the number of accidental coincidences per time bin given by
\begin{align}
	N_a=s_1 s_2 \delta_t T,
\end{align}
where $s_1$, $s_2$ are the count rates of the detectors.

The extracted $g^{(2)}(\tau)$ histograms are fitted with the model function $g^{(2)}(\tau)=1+be^{-2\left|\tau\right|/\tau_c}$ (see Fig. \ref{fig:spectrum and g2}). The fit parameters for the source without amplification (point A) are $b=0.417\pm0.006$ and $\tau_c=4.02\pm0.07\,\text{ns}$. After the first EDFA at point B, we observed $b=0.350\pm0.006$ and $\tau_c=4.30\pm0.09\,\text{ns}$, and after the second EDFA at point C, $b=0.196\pm0.005$ and $\tau_c=4.1\pm0.1\,\text{ns}$. The change in the parameters can be explained by analyzing the output electric field from the EDFAs, which can be modeled as
\begin{align}
	E(t)=E_\text{s}(t)+E_A(t),
\end{align}
where $E_s(t)$ is the amplified light source field and $E_A(t)$ is the ASE field. This modifies $g^{(2)}(\tau)$ to
\begin{align}
	g^{(2)}(\tau)&=\rho g^{(2)}_s(\tau)+(1-\rho)2g^{(2)}_A(\tau)\\&+2\rho(1-\rho)\text{Re}\left[g^{(1)}_s(\tau)g_A^{(1)*}(\tau)\right], \nonumber
\end{align}
where $\rho=\left|E_s(t)\right|^2/\left|E(t)\right|^2$ is the fraction of amplified source light power. As the ASE is thermal light \cite{doronin2019second} with a broadband spectrum, we have $\tau_{c,A}\gg\tau_j$ which washes out the contribution from the second and third term. Hence, the main effect to the extracted $g^{(2)}(\tau)$ is a reduction in peak height $b$. After removing most of the ASE, the peak height increases to $b=0.401\pm0.006$ with $\tau_c=3.98\pm0.08\,\text{ns}$ at point D. This is close to the values at point A with a slight difference caused by residual ASE within the 4\,nm passband of the filter stage. We expect that with a narrower filter passband, these values should match the observations at point A. 

\section{Long range LIDAR with thermal light}
Previous thermal LIDAR experiments were limited to table top demonstrations, optical fiber measurements, or used a retroreflector target with a low return loss \cite{zhu2012thermal,tan2023practical,liu2025active,staffas2025thermal}, as the power levels accessible were too low to overcome the geometrical return loss from diffuse reflection \cite{wandinger2005introduction}. The amplifiers allow us to reach an optical power that is sufficient to overcome these losses.

The setup for the LIDAR demonstration is similar to that used for characterizing the light source, with the inclusion of a pair of identical telescopes (see Fig. \ref{fig:setup}). One telescope is used to illuminate the target, and the other collects the reflected light into a detector. Each telescope consist of an aspheric lens fiber collimator ($f_{1}$=18.75\,mm) and a beam expander, formed by a plano convex lens ($f_{2}$=25.4\,mm) and a second plano convex lens with an aperture of 75\,mm ($f_{3}$=200\,mm). A bandpass filter with bandwidth 4\,nm in the collection telescope removes background light from the environment. Not shown are longpass dichoric mirrors in the beam expanders that inject visible laser light for initial alignment of the telescopes. 

The telescopes are mounted on a pair of stepper motors to provide pitch and yaw control for a LIDAR raster scan.

The measurements were performed at wavelength 1550\,nm for its low atmospheric attenuation \cite{elder1953infrared,giggenbach2022atmospheric} and high "eye-safe" optical power threshold \cite{soltani2022safety}. To prevent complications from sunlight, the measurements were carried out at night. Daylight operation should also be possible with a sufficiently narrow spectral filter, as the source has a narrow spectrum.

The targets are a building and floodlights at two different distances (see Fig. 4). We used an optical power of 50\,mW. This gave a maximum measured return photoevent rate of around $2\times10^5$\,s$^{-1}$ from the diffuse reflection off the building. This  fluctuated between $6\times10^{4}$\,s$^{-1}$ and $2\times10^5$\,s$^{-1}$ over the measurement duration due to difference in surface reflectivity, target distance, and turbulence. An integration time $T=60$\,s per pixel was used, and the $g^{(2)}(\tau)$ was extracted with time bin width $\delta_t=0.25$\,ns.

\begin{figure}
  \centering
  \includegraphics[width=1\columnwidth]{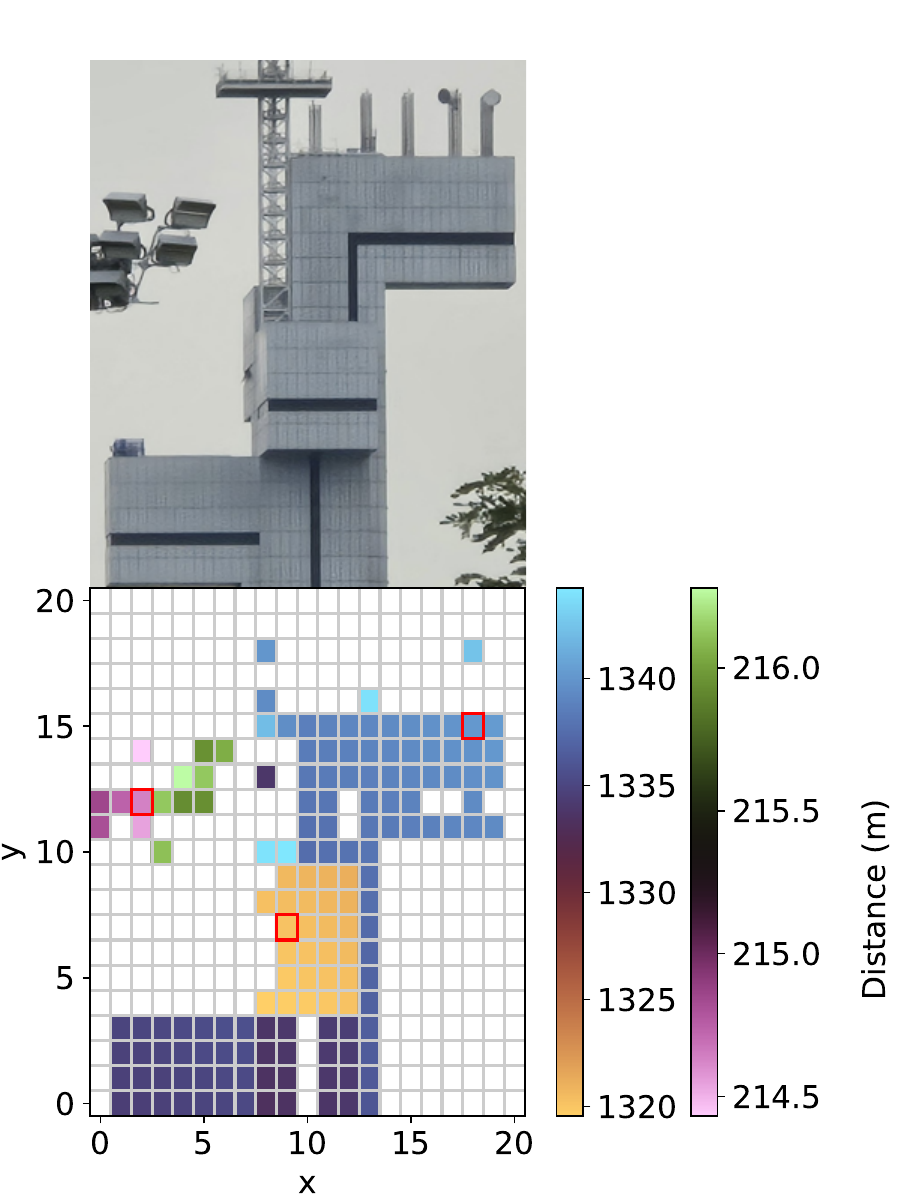}
  \caption{\label{fig:1km distance}
    \textbf{Top:} Photo of the building and floodlights. \textbf{Bottom:} Results of a 21\,pixel x 21\,pixel raster scan. The building is about 1.3\,km away, with lights in the foreground about 215\,m away. White pixels have insufficient return photoevents to identify the peak in $g^{(2)}(\tau)$ with sufficient SNR. 
	}
\end{figure}

\begin{figure}
  \centering
  \includegraphics[width=\columnwidth]{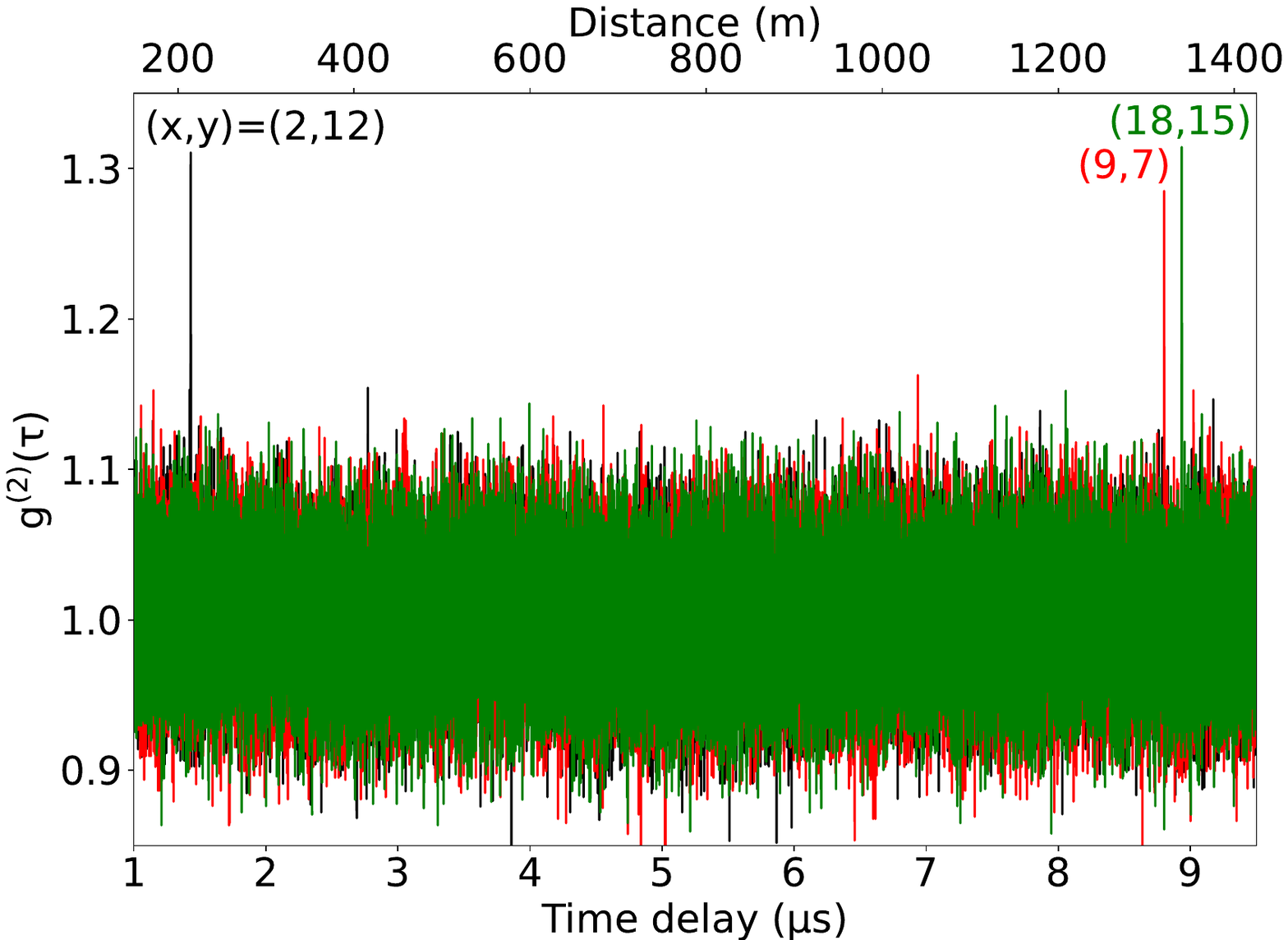}
  
  \caption{\label{fig:1km g2 3 points}
    Extracted $g^{(2)}(\tau)$ for three different points (x, y) highlighted in red in Fig. \ref{fig:1km distance}. (2,12) floodlights, (9,7) middle protrusion of the building and (18,15) top right corner of the building. The $g^{(2)}(\tau_d)$ peaks can be clearly identified at a time delay of 1.43\,\textmu s, 8.80\,\textmu s, and 8.93\,\textmu s, respectively. 
	}
\end{figure}

The additional time difference introduced by the round trip time to the target shifts the $g^{(2)}(\tau)$ peak from $\tau=0$ by $\tau_d=2d/c$, which allows determination of the target distance $d$. Figure \ref{fig:1km distance} shows the result of a 21\,pixel by 21\,pixel raster scan of the targets, and Fig. \ref{fig:1km g2 3 points} shows the extracted $g^{(2)}(\tau)$ for the red highlighted pixels in Fig. \ref{fig:1km distance}. The round trip time for these pixels are $1.43089\pm0.00007\,\text{\textmu s}$, $8.80136\pm0.00007\,\text{\textmu s}$, and $8.93443\pm0.00007\,\text{\textmu s}$ corresponding to distances of $214.49\pm0.01\,\text{m}$, $1319.29\pm0.01\,\text{m}$, and
$1339.24\pm0.01\,\text{m}$, respectively. These are in good agreement with the distances of 215\,m and 1.3\,km for the flood lights and building, that were measured using a commercial laser range finder and google maps. All pixels in Fig. \ref{fig:1km distance} have similar a depth resolution of $\pm0.02\text{m}$. 

This LIDAR measurement is in essence a peak identification problem with signal-to-noise ratio given by
\begin{align}
	\text{SNR}&=\frac{(g^{(2)}(\tau_d)-1)\times N_a}{\sigma_{N_a}}\\
	&=(g^{(2)}(\tau_d)-1)\sqrt{s_1 s_2 \delta_t T}, \nonumber
\end{align}
where $\sigma_{N_a}=\sqrt{N_a}$ arises from Poissonian counting statistics.

A sufficiently large SNR is required for successful peak identification. To achieve this, a comparatively long integration time of 60\,s per pixel was used. The long integration is mainly due to the long dead time of the detectors that limit their maximum count rate to below $10^{6}\,\text{s}^{-1}$. If superconducting nanowire detectors with a maximum count rate in excess of $10^{7}\,\text{s}^{-1}$ \cite{chang2021detecting} were used instead, the same SNR can be achieved in less than a second. Alternatively, by using an analog detector as the reference detector as demonstrated in \cite{jaso2026ghost}, integration times of milliseconds might be possible.

\section{Summary and Outlook}
In this work, we determined $g^{(2)}(\tau)$ of a thermal source before and after optical amplification through stimulated emission, and showed that $g^{(2)}(\tau)$ is preserved after amplification. However, the types of thermal light sources that can be amplified by stimulated emission are rather limited due to the ASE introduced by the amplification. Optical parametric amplification, which has been shown to work at the single photon level, might be one way to overcome this limitation and amplify thermal light sources that are otherwise unsuitable.

The increased optical power and preserved correlations were then used in LIDAR imaging of a building at 1.3\,km away, with depth resolution of 2\,cm. Thus, thermal light can be used for high loss applications like LIDAR, where its intrinsic randomness can be utilized for resilience against crosstalk between different LIDAR units.

\section{Acknowledgments}
This project is supported by the National Research Foundation, Singapore through the National Quantum Office under its Quantum Engineering Programme 3.0 Funding Initiative (W25Q3D0004).

%

\end{document}